\documentclass[10pt,letterpaper,twocolumn]{article}
\usepackage{ol2}
\usepackage{hyperref}
\usepackage{amsmath}
\usepackage{amssymb}
\usepackage{graphicx}
\usepackage[utf8]{inputenc}
\usepackage{ulem}
\usepackage{color}
\usepackage{float}

\graphicspath{{images/}}

\usepackage{color}
\definecolor{gray}{gray}{0.4}
\definecolor{blackcolor}{rgb}{0,0,0}

\hypersetup
{	bookmarksopen=true,
	pdftitle="Compressed Sensing with off-axis frequency-shifting holography",
	pdfauthor="Marcio Marim", 
	pdfmenubar=true, 				
	colorlinks = true, 				
	linkcolor = blue, 				
	citecolor = blue, 			  	
	urlcolor = blackcolor,			
	filecolor = cyan, 				
	pdfpagemode = None, 			
	pdfpagelayout = SinglePage, 	
	pdffitwindow = true 			
}

\begin{document}

\twocolumn[ 

\title{Compressed Sensing with off-axis frequency-shifting holography}

\author{Marcio M. Marim$^{1,3}$, Michael Atlan$^{2}$, Elsa Angelini$^3$ and Jean-Christophe Olivo-Marin$^{1}$}

\affiliation{$^1$Institut Pasteur, Unité d'Analyse d'Images Quantitative, CNRS URA 2582 \\
25-28 rue du Docteur Roux 75015, Paris, France \\
$^2$ CNRS UMR 7587, INSERM U 979, UPMC, UP7, Fondation Pierre-Gilles de Gennes.\\ 
Institut Langevin. ESPCI ParisTech - 10 rue Vauquelin. 75005 Paris. France \\
$^3$Institut Télécom, Télécom ParisTech, CNRS LTCI,
46 rue Barrault 75013, Paris, France \\
Corresponding authors: \href{mailto:marim@pasteur.fr}{marim@pasteur.fr}
}
\date{\today}
\begin{abstract}
This work reveals an experimental microscopy acquisition scheme successfully combining Compressed Sensing (CS) and digital holography in off-axis and frequency-shifting conditions.
CS is a recent data acquisition theory involving signal reconstruction from randomly undersampled measurements, exploiting the fact that most images present some compact structure and redundancy.
We propose a genuine CS-based imaging scheme for sparse gradient images, acquiring a diffraction map of the optical field with holographic microscopy and recovering the signal from as little as 7\% of random measurements.
We report experimental results demonstrating how CS can lead to an elegant and effective way to reconstruct images, opening the door for new microscopy applications.
OCIS : 070.0070, 180.3170
\end{abstract}

] 

\noindent
General high resolution microscopy involves dense data acquisition. One intense field of research aims to reduce the amount of data acquisition or sample illumination \cite{Jackson2009, Hoebe2007}.
In \cite{Jackson2009}, the acquisition is restricted to only those areas where relevant signal is present.
In \cite{Hoebe2007} a method called controlled light-exposure microscopy (CLEM) is introduced, supported by a nonuniform illumination of the field of view.
However, both methods suffer from being image-content dependent for a successful implementation.
Indeed, these methods need a feedback loop inside the acquisition setup to make decisions about the sampling rate or the illumination intensity, depending on the objects characteristics.
Here, we address the sensing problem in microscopy by taking an alternative approach provided by the new theoretical framework of Compressed Sensing (CS).
This method is independent of image-content and does not need any feedback loop during the acquisition.
CS was previously reported in magnetic resonance imaging acquisition \cite{Lustig2007}, single-pixel imaging \cite{Takhar2006} or inline, single-shot holography for tridimensional imaging \cite{Brady2009}.
The main idea presented here is to combine
off-axis, frequency-shifting (for accurate phase-shifting) digital holography to perform quadrature-resolved random measurements of an optical field in a diffraction plane and a sparsity minimization algorithm to reconstruct the image.

CS is a novel mathematical theory for sampling and reconstructing signals in an efficient way, introduced by Candès and Donoho \cite{Candes2006c, Donoho2006, Candes2005}. It exploits the fact that most images are compressible or sparse in some domain due to the homogeneity, compactness and regularity of structures. Instead of sampling the entire data and then compress it to eliminate redundancy, CS performs a compressed data acquisition.
Some basic requirements to enable Compressed Sensing are
(i) to find a sparsifying transform able to shrink the data into a small number of coefficients
(ii) to acquire random projections of the signal into orthogonal subspaces, such as the Fourier domain for spatially-sparse images
(iii) to use a sampling scheme that obeys the Restricted Isometry Property (RIP) \cite{Candes2008a} and
(iv) to use a sampling domain and a sparsifying transform that span incoherent domains 
(i.e. domains where the signal is dense in one case and sparse in the other one)
\cite{Candes2006c}.

Complying with these requirements, CS states that a signal $g \in \mathbb{R}^{N}$ having a $S$-sparse representation (i.e. it can be well represented by a small number $S$ of coefficients, where $S \ll N$) on a basis $\Psi$, can be reconstructed very accurately from a small number of projections of $g$ onto randomly chosen subspaces (e.g. Fourier measurements for spatial sparsity). More precisely, a signal $g$ has a sparse representation if it can be written as a linear combination of a small set of vectors taken from some basis $\Psi$, such as $g = \sum_{i}^{N} c_i \Psi_i$, with ${\parallel c \parallel}_{\ell_1} \approx S$, where ${\parallel \cdot \parallel}_{\ell_1}$ denotes the $\ell_1$ norm which corresponds to the sum of magnitudes of all terms of the candidate signal $g$ projected on $\Psi$. In general, the ${\ell_p}$ norm is defined as ${\parallel \mathbf{c} \parallel}_{\ell_p} := \{ \sum_{i=1}^{N} |c_i|^p \}^{1/p}$.

As demonstrated in \cite{Candes2005}, if such a sparsifying transform $\Psi$ exists in the spatial domain, it is possible to reconstruct an image $g$ from partial knowledge of its Fourier spectrum. In our case, $g$ will represent the local optical intensity in the object plane.  We denote $f \in \mathbb{C}^{N}$ the associated complex optical field, satisfying $g = |f|^2$. The radiation field propagates from the object to the detector plane in Fresnel diffraction conditions.
Thus, the optical field in the object plane $f$ is linked to the field $F$ in the detection plane by a Fresnel transform, expressed in the discrete case as:
\vspace{-2pt}
\begin{eqnarray}
\nonumber
F & = & \mathcal{F}(f) \,  : \, \mathbb{C}^N \rightarrow \mathbb{C}^N\\
F_p & = & \frac{1}{N} \sum_{n=1}^{N} f_n \, {\rm e}^{ i \left( \alpha n^2 - 2 \pi n p /N \right)}
\label{eq_tr_fresnel_discrete}
\end{eqnarray}
where $n, p \in \{1, \dots, N\}$ denote pixel indexes, $\alpha \in \mathbb{R}^+$ is the parameter of the quadratic phase factor ${\rm e}^{ i \alpha n^2}$ describing the curvature in the detection plane of a wave emitted by a point source in the object plane. In CS, the signal reconstruction consists in solving a convex optimization problem that finds the candidate $\hat{g}$ ($\hat{\cdot}$ denotes an estimator) of minimal complexity satisfying $\hat{F}|_{\Gamma} = F|_{\Gamma}$, where $F|_{\Gamma} \subseteq F$ is a partial subset of measurements in the set $\Gamma$.

%
\begin{figure}[!b]
\centering
\includegraphics[width = 8 cm]{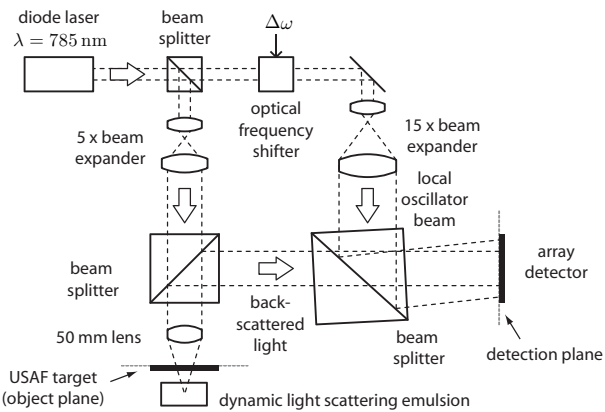}
\caption{Sketch of the experimental image acquisition setup.}\label{fig_080731nanoDLS1}
\vspace{-5pt}
\end{figure}
The experimental setup is sketched in Fig. \ref{fig_080731nanoDLS1}. It consists of an off-axis, frequency-shifting digital holography scheme \cite{Atlan2007a, Gross2007}. The monochromatic optical field from a diode laser dynamically backscattered by an intralipid emulsion illuminates an US Air Force (USAF) resolution target, beats against a separate local oscillator (LO) field detuned by $\Delta \omega / (2 \pi) = 200\, \rm Hz$ and creates a time-fluctuating interference pattern measured with a $N = 1024 \times 1024$ array detector. The diffracted object field map in the detector plane, resolved in quadrature (in amplitude and phase) $F \in \mathbb{C}^N$ is calculated from a four-phase measurement \cite{Atlan2007a}. The frequency detuning $\Delta \omega$ enables rejection of non fluctuating light components reflected by the target as well as speckle reduction through signal accumulation.

$F$ can be back-propagated numerically to the target plane with the standard convolution method when all measurements $F \in \mathbb{C}^N$ are available. In this case, the complex field in the object plane $f$ is retrieved from a discrete inverse Fresnel transform of $F$; $f = \mathcal{F}^{-1}(F)$ :
\begin{equation}\label{eq_tr_fresnel_discrete_inverse}
f_p =  \frac{1}{N} \sum_{n=1}^{N} F_n \, {\rm e}^{-i \left( \alpha n^2 - 2 \pi n p /N \right)}
\end{equation}

\begin{figure}[!h]
\begin{center}
\includegraphics[width=7.5cm]{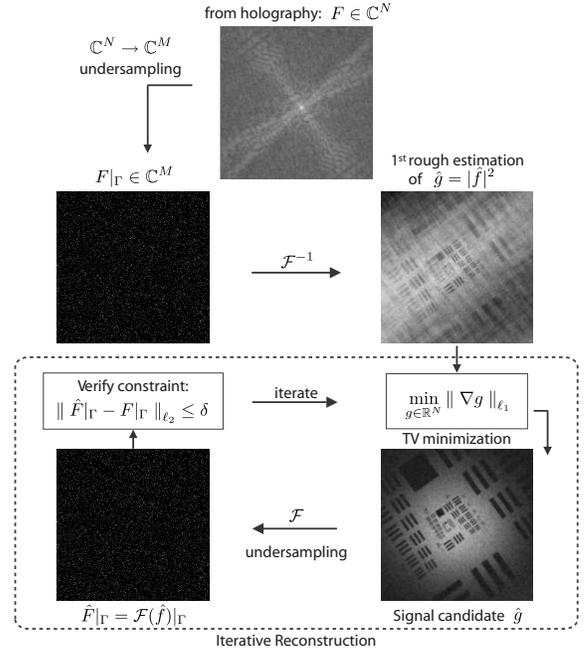}
\end{center}
\vspace{-15pt}
\caption{CS reconstruction scheme.}
\label{fig:scheme}
\end{figure}

Now returning to the CS reconstruction problem, we want to recover the intensity image of the object $g = \{|f|^2 : f \in \mathbb{C}^N \}$ from a small number of measurements $F|_{\Gamma} \in \mathbb{C}^M$ where $M \ll N$. Partial measurements in the detection plane, illustrated by the first step in Fig. \ref{fig:scheme}, can be written as $F|_{\Gamma} = \Phi f$, where the sampling matrix $\Phi$ models a discrete Fresnel transform (eq. \ref{eq_tr_fresnel_discrete}) and random undersampling with flat distribution.
%
To find the best estimator $\hat{g}$, we solve the following convex optimization problem \cite{Candes2006d}:
\begin{eqnarray}
\hat{g} = \arg \min_{g \in \mathbb{R}^N} {\parallel \Psi g \parallel}_{\ell_1}\ \ \mbox{subject to}\ \
\hat{F}|_{\Gamma} = F|_{\Gamma}
\label{eq:optim}
\end{eqnarray}
This optimization leads to an iterative image reconstruction process, illustrated by the loop inside the dotted frame in Fig. \ref{fig:scheme}.
Explicitly, given a partial knowledge of the Fresnel coefficients $F|_{\Gamma}$, we seek a solution $\hat{g}$ with maximum sparsity (i.e. with minimal norm ${\parallel \Psi g \parallel}_{\ell_1}$), and whose Fresnel coefficients $\hat{F}|_{\Gamma}$ match the subset observed $F|_{\Gamma}$ (as illustrated in Fig. \ref{fig:scheme}). 
Since our test image is piecewise constant with sharp edges (such as most microscopy images), it can be sparsely represented computing its gradient. In image processing, a suitable norm to constrain the gradient of an image was introduced as the Total Variation (TV) which measures the $\ell_1$ norm of the gradient magnitudes over the whole image:
\begin{equation*}
{\parallel g \parallel}_{TV} = {\parallel \nabla g \parallel}_{\ell_1}
\end{equation*}
The incoherence property holds for the two basis adopted here, which are the Fresnel spectrum and the TV \cite{Candes2006c}. Moreover, random measurements in the spectral domain satisfy the RIP condition \cite{Candes2008a}. Hence for overwhelming percentage of Fresnel coefficients sets $\Gamma$ with cardinality obeying $|\Gamma| = M \geq K \cdot S \log N$, for some constant $K$, $\hat{g}$ is the unique solution to the problem:
\begin{eqnarray}
\hat{g} = \arg \min_{g \in \mathbb{R}^N} {\parallel \nabla g \parallel}_{\ell_1}\ \ \mbox{subject to}\ \
\hat{F}|_\Gamma = F|_\Gamma
\label{eq:optim2}
\end{eqnarray}
However, holographic measurements are corrupted with noise and the observed signal is not exactly sparse. More appropriately, the observations can be described by noisy measurements $F|_{\Gamma} = \Phi f + n$, where $n \in  \mathbb{C}^M$ is a noise component with bounded energy ${\parallel n \parallel}_{\ell_2} \leq \epsilon$. In this particular case, a better reconstruction can be achieved by relaxing the constraint $\hat{F}|_\Gamma = F|_\Gamma$ and allowing an error $\delta$ at most proportional to the noise energy $\epsilon$ \cite{Marim2009b,Donoho2006b}.
Finally, solving the following problem performs the reconstruction of $g$ with robustness to noise:
%
\begin{equation}
\hat{g} = \arg \min_{g \in \mathbb{R}^N} {\parallel \nabla g \parallel}_{\ell_1}\ \ \mbox{subject to}\ \
{\parallel \hat{F}|_\Gamma - F|_\Gamma \parallel}_{\ell_2}\leq \delta
\label{eq:optim3}
\end{equation}
for some $\delta \leq C \epsilon$, which depends on the noise energy.

%
\begin{figure}[!b]
\begin{center}
\includegraphics[width=8.4 cm]{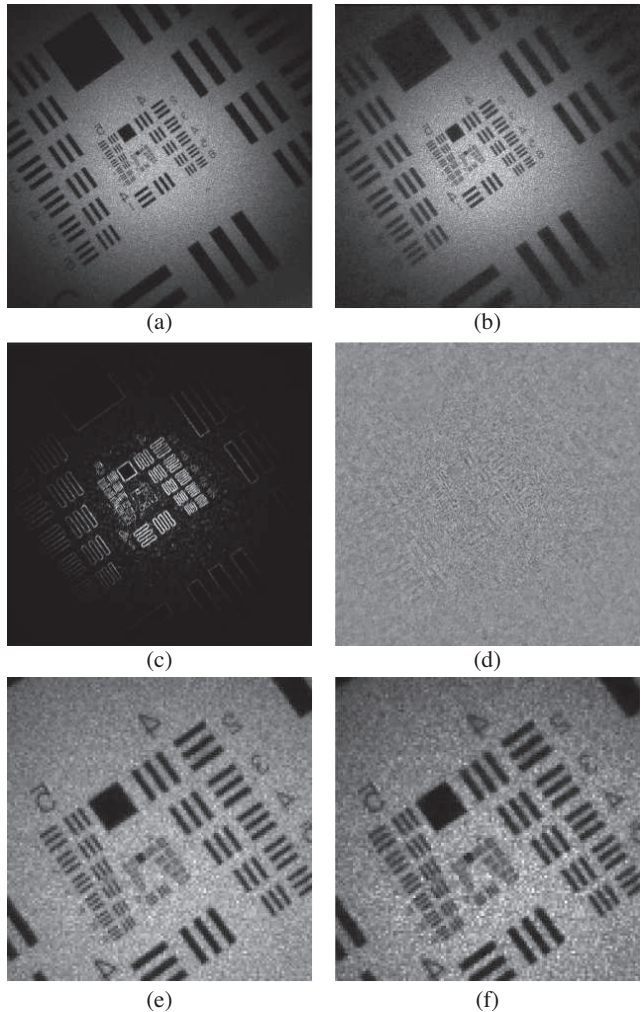}
\end{center}
\vspace{-15pt}
\caption{(a) Standard holography, as described in Eq. (\ref{eq_tr_fresnel_discrete}). (b) CS reconstruction, using 7\% of the Fresnel coefficients. (c) Gradient of $g$. (d) Residual from (a) and (b). (e), (f) Magnified views from (a) and (b).}
\label{fig:result}
\vspace{-10pt}
\end{figure}
In Fig. \ref{fig:result} we illustrate some CS reconstruction results. A reconstruction of an off-axis image with the standard convolution method (eq. \ref{eq_tr_fresnel_discrete_inverse}) is illustrated in Fig. \ref{fig:result}a. The image reconstructed with holography uses all available measurements (4 phases $\times$ 10 accumulations $\times 1024^2 = 4.2 \times 10^7$ pixels).
For the CS approach, Fresnel coefficients are undersampled randomly. Fig. \ref{fig:result}b shows the CS reconstruction result from only 7\% of the pixels used in the standard approach (4 phases $\times$ 10 accumulations $\times$ 0.07 $\times 1024^2 = 2.9 \times 10^6$ pixels).
Fig. \ref{fig:result}c illustrates the gradient of the image $\nabla g$ (sparsifying domain) and Fig. \ref{fig:result}d illustrates the residual (Euclidean distance $|\hat{g}-g|$) from standard holographic reconstruction (a) and CS reconstruction (b). The global normalized error is ${\parallel \hat{g}-g \parallel}_{\ell_2} = 0.005 $ ($\hat{g}$ and $g$ have unit norms).
This error is essentially due to the relaxation of the constraint for a perfect match between measures and estimations in the CS scheme, leading to some denoising effect, confirmed by the visual aspect of the residual image image Fig. \ref{fig:result}d showing essentially unstructured noise.
Finally, Figs. \ref{fig:result}e and \ref{fig:result}f display magnified views from central region of images (a) and (b), illustrating the quality of the reconstruction.
%

\bigskip
In conclusion, we have presented a novel microscopy imaging framework successfully employing Compressed Sensing principles.
It combines an iterative image reconstruction
and digital holography to perform quadrature-resolved random measurements of an optical field in a diffraction plane.
The CS approach enables optimal image reconstruction while being robust to high noise levels.
The proposed technique is expected to greatly improve many microscopy applications, allowing the acquisition of high dimensional data with reduced acquisition time increasing imaging throughput and opening the door to sample-friendly acquisition protocols.

\bigskip
This work was funded by Institut Pasteur, Direction Générale de l'Armement (DGA), Institut Langevin, ANR and CNRS. The authors also acknowledge support from Fondation Pierre-Gilles de Gennes.



\end{document}